\documentclass[reprint,superscriptaddress,
pra,
amsmath,amssymb]{revtex4-1}
\usepackage{graphicx}
\usepackage{dcolumn}
\usepackage{sidecap}
\usepackage{bm}
\usepackage{color}
\usepackage{textgreek}
\usepackage{hyperref}
\usepackage{amsmath}
\usepackage{titlesec} 
\titlespacing\section{0pt}{10pt}{4pt}
\titlespacing\subsection{0pt}{10pt}{2pt}

\begin{document}
\title{Probing topological spin structures using light-polarization and magnetic microscopy}
\author{Till Lenz}
\affiliation{Helmholtz Institut Mainz, Johannes Gutenberg-Universit\"at Mainz, 55128 Mainz, Germany}
\author{Georgios Chatzidrosos}
\affiliation{Helmholtz Institut Mainz, Johannes Gutenberg-Universit\"at Mainz, 55128 Mainz, Germany}
\author{Zhiyuan Wang}
\affiliation{Helmholtz Institut Mainz, Johannes Gutenberg-Universit\"at Mainz, 55128 Mainz, Germany}
\author{Lykourgos Bougas}\email{lybougas@uni-mainz.de}
\affiliation{Institut f\"ur Physik, Johannes Gutenberg-Universit\"at Mainz, 55128 Mainz, Germany}
\author{Yannick Dumeige}
\affiliation{Universit\'e de Rennes, CNRS, Institut FOTON UMR 6082, F-22305 Lannion, France}
\author{Arne Wickenbrock}
\affiliation{Helmholtz Institut Mainz, Johannes Gutenberg-Universit\"at Mainz, 55128 Mainz, Germany}
\author{Nico Kerber}
\affiliation{Institut f\"ur Physik, Johannes Gutenberg-Universit\"at Mainz, 55128 Mainz, Germany}
\author{Jakub Z\'azvorka}
\affiliation{Institut f\"ur Physik, Johannes Gutenberg-Universit\"at Mainz, 55128 Mainz, Germany}
\affiliation{Institute of Physics, Faculty of Mathematics and Physics, Charles University, 121 16 Prague, Czech Republic}
\author{Fabian Kammerbauer}
\affiliation{Institut f\"ur Physik, Johannes Gutenberg-Universit\"at Mainz, 55128 Mainz, Germany}
\author{Mathias Kl\"aui}
\affiliation{Institut f\"ur Physik, Johannes Gutenberg-Universit\"at Mainz, 55128 Mainz, Germany}
\author{Zeeshawn Kazi}
\affiliation{University of Washington, Physics Department, Seattle, WA, 98105, USA}%
\author{Kai-Mei C. Fu}
\affiliation{University of Washington, Physics Department, Seattle, WA, 98105, USA}%
\affiliation{University of Washington, Electrical and Computer Engineering Department, Seattle, WA, 98105, USA}
\affiliation{Helmholtz Institut Mainz, Johannes Gutenberg University, 55128 Mainz, Germany}
\author{Kohei Itoh}
\affiliation{Spintronics Research Center, Keio University, 3-14-1 Hiyoshi, Kohoku-ku, Yokohama 223-8522, Japan}
\author{Hideyuki Watanabe}
\affiliation{Device Technology Research Institute, National Institute of Advanced Industrial Science and Technology,Tsukuba Central 2, 1-1-1 Umezono, Tsukuba, Ibaraki 305-8568, Japan}
\author{Dmitry Budker}\email{budker@uni-mainz.de}
\affiliation{Helmholtz Institut Mainz, Johannes Gutenberg-Universit\"at Mainz, 55128 Mainz, Germany}
\affiliation{Department of Physics, University of California, Berkeley, California 94720-300, USA}
\date{\today}

\begin{abstract}
We present an imaging modality that enables detection of  magnetic moments and their resulting stray magnetic fields. We use wide-field magnetic imaging that employs a diamond-based magnetometer and has combined magneto-optic detection (e.g. magneto-optic Kerr effect) capabilities. We employ such an instrument to image magnetic (stripe) domains in multilayered ferromagnetic structures. 
\end{abstract}
\maketitle

\section{Introduction}
Understanding the behaviour of spins and charges in magnetic systems is at the heart of condensed matter physics, and intense ongoing research activities are focused on developing and understanding these, towards the generation of faster, smaller, and more energy-efficient magnetic technologies\,\cite{Stamps2014,Sander2017}.\\
\indent The success of all research activities relies on advances in theory and materials synthesis, but most critically on sensitive probes, since an improved understanding of the interplay between the spin-spin and spin-orbit interactions in these systems is realized through the determination of their static and dynamic spin configurations and current distributions. To achieve this, several powerful techniques for real-space probing of magnetic structures are employed, such as spin-polarized low-energy electron microscopy\,\cite{Gabaly2006,Rougemaille2010} and X-ray magnetic circular dichroism\,\cite{Fischer2017,Bonetti2017}, both of which are sensitive to the magnetization distribution of the structures.\\
\indent An alternative approach is the detection of the stray magnetic fields generated by the magnetic textures. Techniques such as magnetic resonance force microscopy\,\cite{Slides1995,Lee2010} and scanning magnetometry using superconducting quantum interference devices\,\cite{Kirtley1999,Vasyukov2013}, allow for real-space imaging of the stray magnetic fields emanating from magnetic structures. However, these techniques typically operate over a narrow range of environmental conditions, and in some cases can have magnetic (perturbative) back-action on the devices under investigation.\\
\indent Magnetometry based on the electron spin of nitrogen-vacancy (NV) defects in diamond has emerged as versatile, highly sensitive stray-field probe for the non-invasive study of magnetic systems\,\cite{Casola2018}. Diamond-based magnetometers can operate from cryogenic to above room temperature environments, have a dynamic range spanning at least nine decades (DC-GHz), allow for sensor/sample distances as small as a few nanometers, and provide access to static and dynamic magnetic/electronic phenomena with diffraction-limited to nanoscale spatial resolution depending on the imaging modality. Most crucially, NV-based magnetometry is magnetically non-perturbative and works under a wide range of external magnetic and electrical fields.\\
\indent Scanning NV-based magnetic microscopy, in which a diamond nanocrystal that hosts a single NV-center is attached to the tip of an atomic force microscope, has already been successfully applied in studies of novel emerging magnetic phenomena, such as in the case of nanoscale imaging and control of domain-wall hopping in ultrathin ferromagnets\,\cite{Tetienne2015} and magnetic wires\,\cite{Tetienne1366}, imaging of non-collinear antiferromagnetic order in magnetic thin films\,\cite{Gross2017}, the direct measurement of interfacial Dzyaloshinskii-Moriya interaction in ferromagnetic multilayer heterostructures\,\cite{Gross2016}, and even in identifying the morphology of isolated skyrmions in ultrathin magnetic films\,\cite{Gross2018,Jenkins2019}. Scanning NV-based magnetometry is highly sensitive and produces magnetic field maps of a sample with spatial resolution ultimately limited by the atomic size of the tip. However, similarly to magnetic force microscopy, its operation requires specific environmental conditions and is not suitable for wide-field inspection of dynamics at fast time scales ($\sim$ms). An alternative is wide-field NV-based magnetic microscopy that employs ensembles of NV centers within a diamond crystal, and allows for magnetic imaging with an unprecedented combination of temporal (ms) and spatial (diffraction-limited) resolution. State-of-the-art experimental demonstrations have demonstrated sub-\textmu T \textmu m$/\sqrt{\rm{Hz}}$ magnetometric sensitivities over wide fields-of-view (FOV) (as large as $\sim$1\,mm$^2$)\,\cite{Pham2011,Chipaux2015,Glenn2015,Fescenko2019}. \\
\indent An additional merit in using NV-based magnetometry is the possibility to simultaneous measurement of all Cartesian components of static/dynamic magnetic fields\,\cite{Pham2011,Schloss2018}. This becomes vital when one wishes to identify the type and the chirality of domain walls of skyrmions in magnetic multilayer stacks\,\cite{Tetienne2015,Dovzhenko2018}. However, even with the employment of vector magnetometric protocols, reconstruction of the magnetic spin-structure topology based on the detected stray magnetic fields is an under-constrained inverse problem (i.e. an infinite number of magnetic topologies can give rise to similar stray magnetic-field patterns)\,\cite{Meltzer2017,Dovzhenko2018,Casola2018}. 
Therefore, a straightforward solution is the implementation of a detection modality that enables the detection of both the magnetization and its resulting stray magnetic fields. \\
\indent We present here a novel imaging modality that allows for the detection of both magnetic moments and their resulting stray magnetic fields, which consists of a polarization-sensitive epifluorescent microscope that incorporates a diamond sensor and exploits the radiation required for the magnetometric measurements to perform magneto-optic Kerr effect (MOKE) measurements\,\cite{McCord2015}. Such an instrument allows for  magnetization and magnetic field detection, while being magnetically non-perturbative and operable over a broad temperature range. Note here, that pertubations might occur due to high laser powers and the resulting heating. However, when using low laser power, the above-mentioned attributes make this sensing approach ideal for the study of novel magnetic structures and their dynamics under a wide range of environmental conditions.

\section{Principles of magneto-optical and wide-field magnetic imaging}

\subsection{NV-based magnetic imaging}
The NV center in diamond consists of a substitutional nitrogen atom in the carbon lattice adjacent to a vacancy. The negatively charged NV center (NV$^{-}$) has an electronic spin-triplet ground state ($^{3}$A$_{2}$; $S = 1$) with magnetic Zeeman sublevels $m_s = \{ -1, 0, +1\}$ (quantized along the N-V binding axis) and an axial zero-field splitting of $\approx$2.87\,GHz between the $m_s = 0$ and $m_s = \pm 1$ sublevels, whose energies shift in response to local magnetic fields, crystal stress, temperature changes, and electric fields (Fig.\,\ref{fig:fig1}\,a). The spin state of the NV-center can be initialized (polarized to the $m_s = 0$ sublevel) using (continuous or pulsed) optical excitation (typically with $\lambda=532$\,nm) and readout through spin-dependent photoluminescence (PL): the $m_s = 0$ sublevel exhibits a higher fluorescence rate under illumination than the $m_s = \pm1$ sublevels. Applying a small external magnetic field along the NV quantization axis lifts the degeneracy of the $m_s = \pm 1$ magnetic sub-levels, and magnetic-field measurements are consequently possible via the precise measurement of the NV spin-resonance frequencies. This is typically done by sweeping the frequency of an externally applied microwave (MW) driving field while monitoring the spin-dependent photoluminescence, thus allowing, for NV-fluorescence-based magnetometry by optically detected magnetic resonance (ODMR) (there exist also alternative magnetometric modalities that rely on absorption- or photocurrent-based detection schemes\,\cite{Chatzidrosos2017,Bourgeois2015}). In high quality diamond sensors, the attainable ODMR linewidth is narrow enough to observe the NV-center's hyperfine structure, i.e. the coupling between the host nitrogen nucleus (99.6\% of $^{14}$N in natural abundance, for which $I_N=1$) and the unpaired NV electron spin [for the case of $^{14}$N, every observed ODMR peak splits into a triplet with a separation of $A_{hfs}\approx2.16$\,MHz\,\cite{Felton2009} (see Fig.\,1\,a inset)].\\
\indent Using ensembles of NV-centers, it is possible to perform highly sensitive magnetic imaging using wide-field (e.g., epi-fluorescence) optical microscopy. In such a case, a thin layer of NV centers is engineered close to the surface of a diamond crystal and is used for two-dimensional mapping of the magnetic field generated by a sample proximal to its surface, by projecting the NV-fluorescence onto a camera sensor and performing ODMR measurements\,\cite{Steinert2010}. Sub-micrometre resolved magnetic-field maps can be achieved using high magnification and high numerical-aperture (NA) microscopes equipped with a common charged-coupled device (CCD)/complementary  metal-oxide-semiconductor (CMOS) camera. Furthermore, NV centers are oriented along one of four crystallographic $\langle 111\rangle$ directions in diamond, and a diamond crystal hosting an NV ensemble typically contains an equal number of NVs for each direction. As such, ODMR spectra from all NV orientations yield the necessary information to reconstruct vector magnetic field components from magnetized structures proximal to the NV centers. Wide-field NV-based imaging techniques have already been used to measure magnetic fields generated by magnetic thin films\,\cite{Simpson2016} and nanoparticles\,\cite{Sage2013}, vortices in type-II superconductors\,\cite{Schlussel2018}, and even to reconstruct the current flow inside integrated circuits\,\cite{Nowodzinski2015} or graphene field-effect transistors\,\cite{Lillie2019}.

\subsection{Magneto-optical imaging}
\begin{figure}[ht]
\centering
\includegraphics[width=\columnwidth]{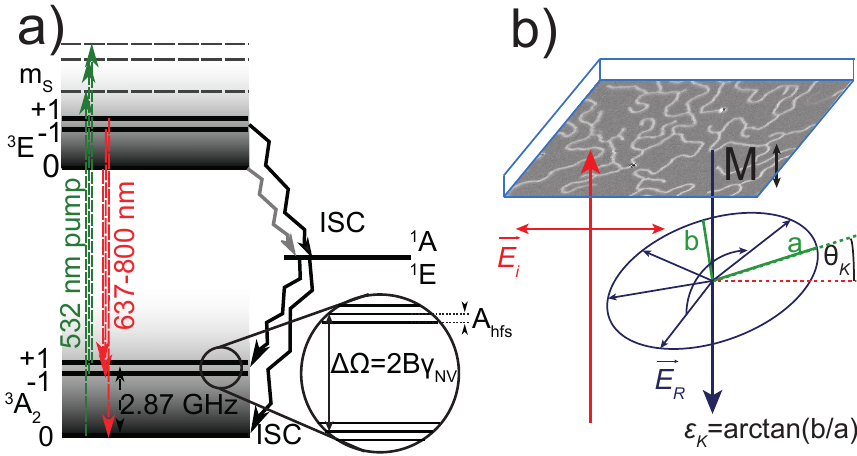}
\caption{\small{(a) NV-energy-level diagram
(b) Schematic of the polar MOKE. $\vec{E_i}$ describes the linear polarized incident light and $\vec{E_R}$ the reflected light with rotation angle $\Theta_{K}$ and ellipticity $\varepsilon_{K}$. } }
\label{fig:fig1}
\end{figure}

Magneto-optical effects from magnetized media can be observed via transmission- or reflection-based measurements. The optical effects that manifest themselves when light is reflected from the surface of a magnetized material are conventionally designated as MOKE, and MOKE microscopy, i.e. a technique capable of studying MOKE with the use of an optical microscope, is one of the most prominent tools for the spatio-temporal visualization of distributions of magnetization (M) within magnetic materials\,\cite{McCord2015}. In general, Kerr effects manifest as changes of the intensity and/or the plane of polarization ($\Theta_{K}$) and the ellipticity ($\varepsilon_{K}$) of an incident  linearly polarized light upon reflection, and all different types of Kerr effects (polar, longitudinal, transverse) scale linearly with magnetization. As such, investigation of the magnetization of a magnetic sample becomes possible with the use of a simple optical microscope that consists of a polarization selection and a polarization analysis step. For most cases of interest, as in our work here, the polar type of MOKE is typically studied, which means that one senses the magnetization component perpendicular to the sample's surface (Fig.\,\ref{fig:fig1}\,b). Moreover, the polar MOKE only manifests as the change of polarization angle ($\Theta_{K}$) and ellipticity ($\varepsilon_{K}$), while the reflected intensity remains unchanged.

\subsection{Concurrent MOKE/magnetic imaging microscopy}
Combination of both imaging modalities, NV-based magnetic and MOKE, is possible by the additional incorporation of polarization-sensitive preparation and analysis steps in the optical setup required to perform NV-based magnetic imaging. In particular, by pre-selecting the polarization state of the optical excitation required for the NV-based magnetometric protocol (i.e., green laser light), and imaging changes in its polarization state upon reflection from a magnetized sample placed upon the diamond crystal, while simultaneously imaging the NV-fluorescence, one obtains a concurrent image of magnetization and their resulting magnetic fields.

\section{Experimental Methods}  

\begin{figure*}[ht]
\centering
\includegraphics[width=\textwidth]{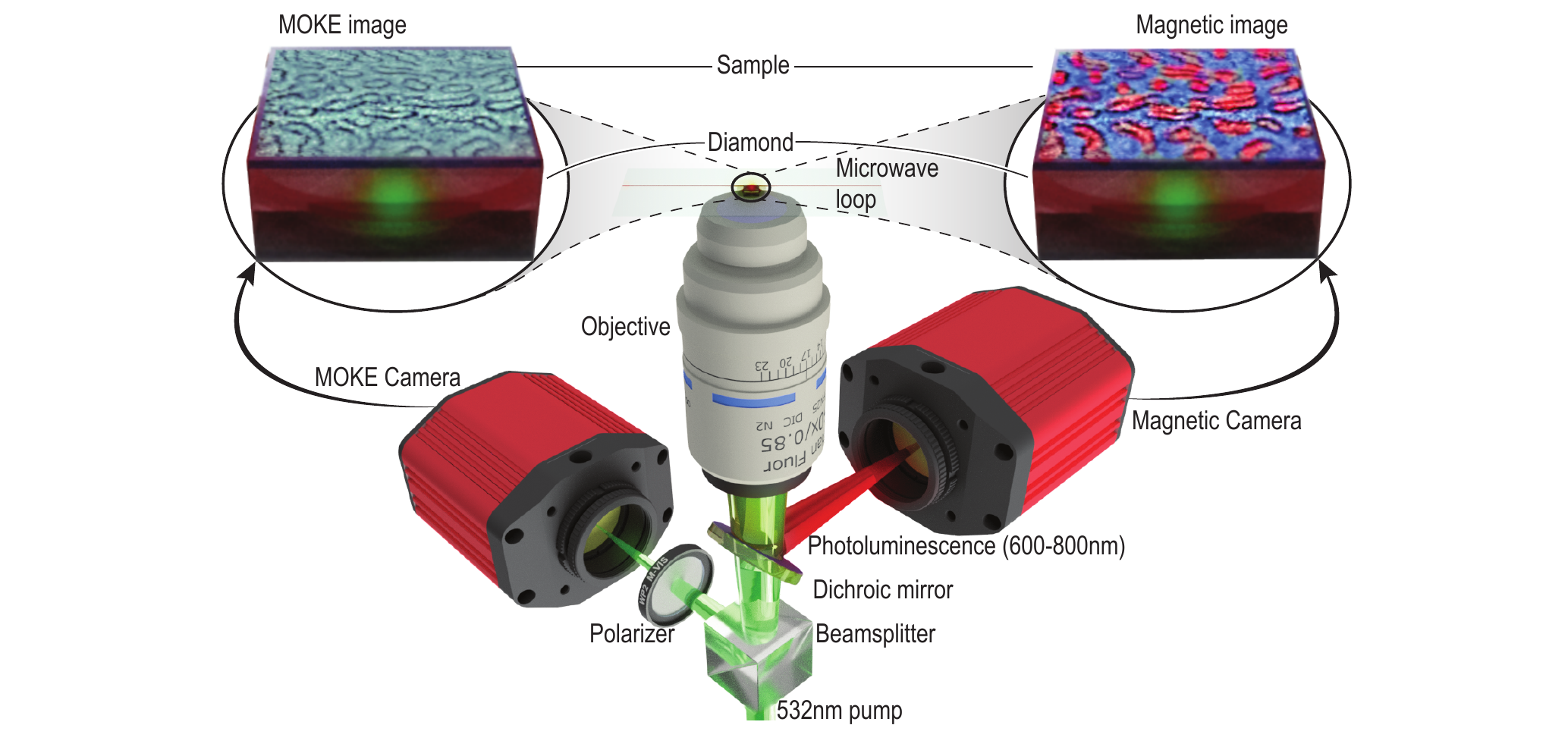}
\caption{\small{Schematic of the experimental setup.}}
\label{fig:fig2}
\end{figure*}

\subsection{Magnetic sample}
To demonstrate the capabilities of an imaging setup that can perform measurements of magnetization and magnetic fields, we use a multilayered ferromagnetic structure that shows rich topological magnetic structures (magnetic domains) under the influence of external magnetic fields.\\
\indent The sample is a Ta(5\,nm)/Co$_{20}$Fe$_{60}$B$_{20}$(1\,nm)/ Ta(0.08\,nm)/MgO(2\,nm)/Ta(5\,nm) material stack that is grown on a Si/SiO$_2$ substrate (500\,\textmu m thickness) by DC magnetron sputtering (using a Singulus Rotaris sputtering system allows us to tune the thickness of the layers with high accuracy; reproducibility better than 0.01\,nm)\,\cite{Zazvorka2019}. The Ta(0.08\,nm) insertion layer, that is not even a complete Ta monolayer, is inserted to fine tune the perpendicular (out-of-plane) magnetic anisotropy stemming from the Co$_{20}$Fe$_{60}$B$_{20}$/MgO interface\,\cite{Yu2016}, to allow us to get closer to the spin reorientation transition\,\cite{Miao2013}. Thereby, we can achieve a typical hour-glass shaped hysteresis loop indicating the presence of magnetic domains or a skyrmion phase\,\cite{Lemesh2018}. Using superconducting quantum interference device (SQUID)-based measurements, we determine the saturation magnetization of the sample to be $M_s \approx 760$\,kA/m at room temperature.

\subsection{Experimental setup}
\textit{\textbf{Optical Setup -} } In Fig.\,\ref{fig:fig2}, we present the experimental setup for the concurrent magnetic field (NV) and magnetization (MOKE) measurements. The basis of the setup is a home-built inverted epifluorescence microscope. \\
\indent As light source for the magnetometric protocol and for the light-polarization measurements, we use a 532\,nm diode laser (Laser Quantum, Gem 532\,nm), which is intensity-stabilized with a feedback system using an acousto-optical modulator (AOM, ISOMET-1260C with an ISOMET\,630C-350 driver) controlled with a proportional-integral-derivative controller (PID, SRS SIM960). The excitation and reflected/PL beams are split using a 50:50 beam splitter cube (BSC) before the objective. To operate in wide-field configuration another lens (f$=150$\,mm) is placed before BSC which leads to an enlarged excitation area.\\
\indent The imaging system consists of an air objective (Olympus UplanFL N 60X) with magnification of 60$\times$, a working distance of 0.2\,mm, and a correction collar which allows us to compensate for cover glass thicknesses of 0.11-0.23\,mm. As detector we use a scientific-CMOS (sCMOS) camera (Andor Zyla\,5.5) with a pixel area of $6.5\times6.5\,$ \textmu$\textrm{m}^{2}$. This leads to an effective pixel area of $\approx108\times108$\,nm$^{2}$ in the focal plane, i.e. on the sample. The home-built sample holder consists of a thickness \#0 ($\approx85-115\,$\textmu m) coverslip which is glued onto a printed circuit-board that is supported upon a 3D translation stage. The coverslip is equipped with an omega shaped stripline to deliver the MW fields required for NV-based magnetometry. On top of coverslip we place our diamond sensor, and on top of the diamond sensor the sample under investigation. To fix the sample-sensor distance distances ($<10$\,\textmu m) and to reduce the amount of aberrations, we use microscope immersion oil between the coverslip/diamond and diamond/sample surfaces. \\
\indent To image the NV PL, we use a 650\,nm longpass filter (FEL0650, Thorlabs) to remove the reflected green light. For MOKE imaging, we remove the longpass filter and introduce a neutral density filter (optical density$=3$) before the camera to coarsely match the amount of reflected photons from the green laser beam to the number of detected photons in the observed NV fluorescence (under typical experimental conditions) for ease of operation (i.e. use similar acquisition settings) while sacrificing optimal polarimetric sensitivities. 

\textit{\textbf{Magnetic Sensor -} } For wide-field NV magnetic imaging we use a near-surface, high density NV ensemble. In particular, a 100\,nm $^{14}$N doped, isotope-purified (99.9$\%$ $^{12}$C) layer was grown by chemical vapor deposition on an electronic-grade diamond substrate (Element Six). The sample was then implanted with 25\,keV He$^{+}$ at a dose of $10^{12}$\,ions/cm$^{2}$ to form vacancies, followed by a vacuum anneal at 900\,$^{\circ}$C for 2\,hours for NV formation and an anneal in $\text{O}_{2}$ at 425\,$^{\circ}$C for 2\,hours for charge-state stabilization\,\cite{Kleinsasser2016}. The resulting ensemble has an NV density of $\sim1.2\times10^{17}$~cm$^{-3}$. 

\textit{\textbf{External fields -} } To generate the magnetic domains in the sample, we apply a bias magnetic field (B$_{z}$) with the use of a coil (length 2.8\,cm, radius 3\,cm, producing approximately $\sim$6\,G/A). The coil is supplied with DC current using a computer controlled power supply.\\
\indent To perform the NV-based magnetic imaging, a MW source (SRS SG384) is used. The MW signal is split and mixed with a 2.16\,MHz signal and then combined again, in order to resonantly address and drive all (three) hyperfine-resolved NV spin resonances simultaneously. This allows us to obtain full contrast ODMR spectra without the the drawback of power broadening and, thus, yields improved mangetometric sensitivities\,\cite{Barry14133,Kazi2020}. All MW frequency components are  amplified with a 16\,W (+43dB) amplifier (ZHL-16W-43+) and passed through a circulator (Pasternack CS-3.000) and a high-pass filter (Mini Circuits VHP-9R5), before they are transmitted through the omega shaped stripline. \\
\indent \textit{\textbf{Polarization control -} } To perform MOKE imaging in polar configuration the setup is equipped with two linear polarizers: one is placed in front of the BSC to prepare the green light with linear polarization (it also enables for optimization of the magnetometric protocol through the appropriate selection of polarization), and the second one, hereafter called the analyzer, is used for analysis of the polarization state of the reflected light beam and is placed directly in front of the camera. The rotation angle of the analyzer with respect to the polarizer was optimized to obtain the best signal-to-noise ratio (SNR) for MOKE measurements. For the present experimental conditions, this optimum was found when the analyzer was nearly crossed with the polarizer.
\indent To allow for switching between magnetic field and magnetization imaging, a computer controlled flip mount was positioned in front of the analyzer that allowed us to switch from using the longpass filter (and thus performing magnetic field measurements) to using the neutral density filter (allowing for MOKE measurements). 

\subsection{Magnetic imaging}
Obtaining quantitative images of the local magnetic field requires acquisition of ODMR spectra for all pixels within the illuminated FOV. Here we employ a CW technique for NV ODMR-based magnetic imaging, wherein optical NV spin polarization, MW drive, and spin-state readout via the NV PL occur simultaneously. The diamond sensor is continuously illuminated with $\approx80$\,mW of 532\,nm laser light, and the resulting NV PL is collected with the objective.\\
\indent For each frequency point of the NV ODMR spectrum ($\nu_{\textrm{i}}$), PL images are captured and normalized by images captured at a fixed off-resonant frequency (in our case $\nu_{\textrm{off}}=2.16$\,GHz). The number of detected photons on each pixel at $\nu_{\textrm{i}}$ is then divided by the detected photon number at $\nu_{\textrm{off}}$ for the same pixel. This sequence is then repeated between 15 and 40 times (for every selected $\nu_{\textrm{i}}$); a typical exposure time is 100\,ms per sequence.\\
\indent Since we are imaging fields from a sample with a perpendicular (out-of-plane) magnetic anisotropy, and the bias magnetic field required to manipulate this anisotropy is applied along the [100] axis of the diamond crystal, the applied field and the resulting magnetic fields from the magnetic structure lead to (almost) equal magnetic field projections along all (eight) different NV orientations (along four axes) in our diamond sample. This, in turn, results in degenerate magnetic eigenstates and, thus, equal transition frequencies in addressing NVs along the different axes. In fact, the magnetic stray fields generated by the sample are not exactly oriented along the applied field, but since the splitting created by transverse components is smaller than the linewidth this results only in broadening of the resonances. As mentioned earlier, due to hyperfine interactions this results in two sets of three transition frequencies, which are separated by $2\gamma_{NV}$B$_{NV}$ ($\gamma_{NV}\approx28$ MHz/mT and B$_{NV}$ projection of the magnetic field along the NV axis). In order to increase the sensitivity of our sensor, we simultaneously apply three frequency components with a fixed separation of 2.16\,MHz matching the hyperfine splitting of the resonances. This frequency triplet is then scanned over the three NV resonances which results in five dips for both $\Delta m_s=\pm1$ transition families. The relative amplitude of the five lorentzians is 1:2:3:2:1 corresponding to the number of applied frequencies which are in resonance with a transition (see Fig.\ref{nvmoke}c and \cite{Barry14133}).  As a result, the observed features in the obtained ODMR spectra are fitted with two sets of five lorentzians (in the general case of an arbitrarily oriented magnetic field eight sets of five lorentzians are observed), and by determining their frequency separation we can image the local magnetic field perpendicular to the diamond surface. We determine the average magnetometric sensitivity of our current setup  to be $\delta \rm{B}_{\rm{NV}} \approx2\,$\textmu T \textmu m/$\sqrt{\rm{Hz}}$ within a FOV of approximately $40\times40\,$\textmu m$^2$, which is mainly limited by mechanical instabilities of the setup.


\subsection{MOKE imaging}
MOKE images of the sample-magnetization topology are taken in a similar fashion to the magnetic images after switching to MOKE operation by using the flip mount to allow for the collection of the green laser light instead of the NV PL.\\
\indent For a given strength of the externally applied magnetic field required to change the magnetization state of the sample, we obtain a MOKE signal for each pixel by averaging over 50 acquisition cycles of the camera (typical exposure time: 10\,ms per cycle). This average is then normalized to an equivalent measurement performed at a reference field $\vert B_{ref}\vert >\vert B_{\textrm{sat}}\vert$ [i.e., $S_{MOKE}=[N_{\textrm{ph}}(B_{\textrm{z}})-N_{\textrm{ph}}(B_{\textrm{ref}})]/N_{\textrm{ph}}(B_{\textrm{ref}})$]. In our case, $B_{\textrm{sat}}$, the saturation field at which the magnetic sample is in a monodomain state with all the domains/spins aligned with the external magnetic field, is approximately $B_{sat}\approx1$\,mT. We determine the polarimetric angle-imaging-sensitivity of our current setup to be $\delta \Theta_{K}\approx50\,$\textmu rad\,\textmu m/$\sqrt{\rm{Hz}}$ (again mainly limited by mechanical instabilities of the setup).


\section{Results}

\begin{figure*}
\includegraphics[width=0.9\textwidth]{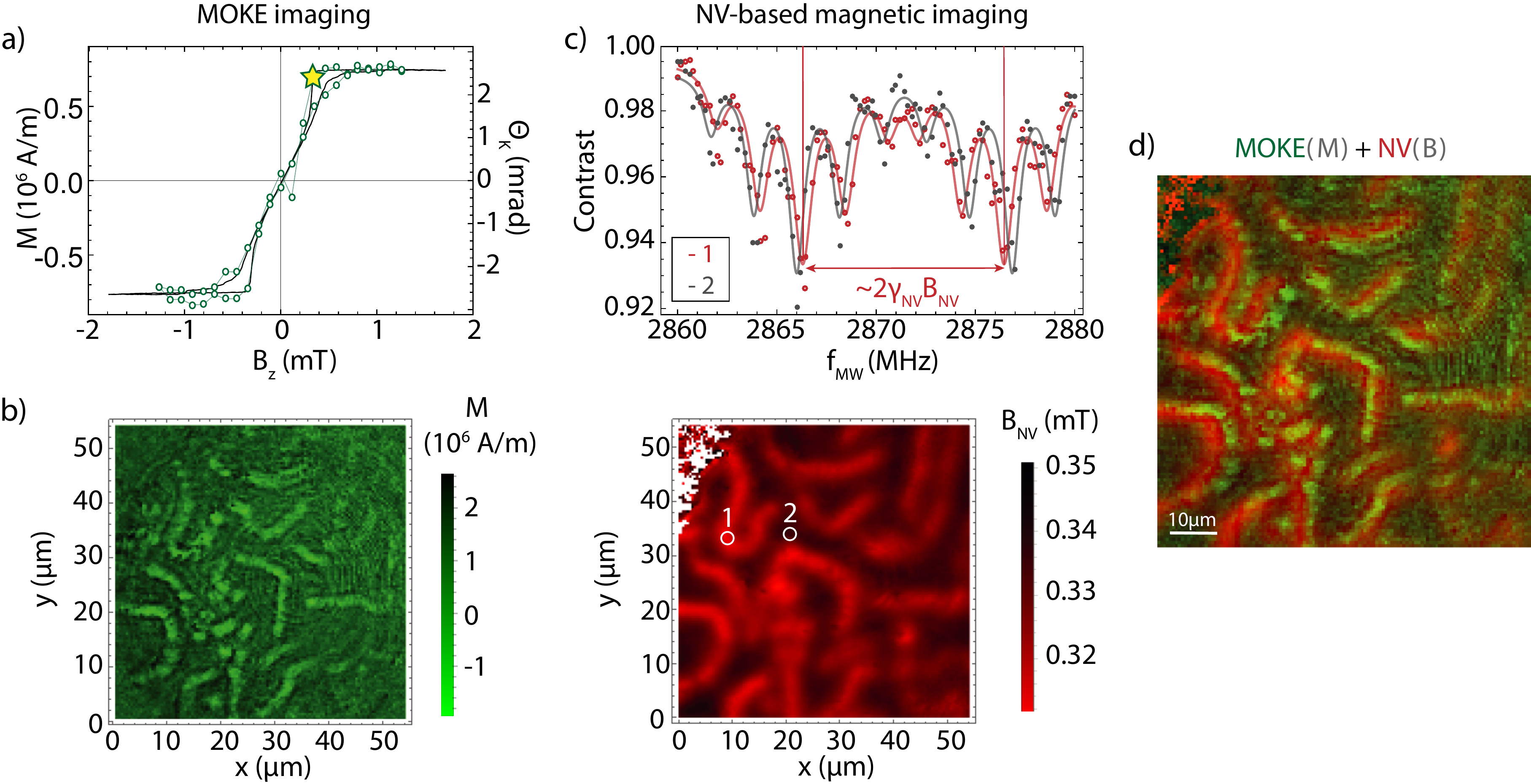}
\caption{\small{\textit{Magnetization and magnetic field imaging using MOKE and NV-based magnetic field imaging -} (a) Out-of-plane magnetization curve of the magnetic sample (open green circles), and for comparison we present the magnetization curve obtained using a commercial MOKE microscope (black line). The starred point denotes the position in the hysteresis curve where the MOKE and NV-based magnetic images were obtained. (b) MOKE image for an applied external magnetic field of B$_z=0.316$\,mT. (c) ODMR spectra for two different positions within the imaged field-of-view (points are measured data while the lines represent the best fit for these data sets) and imaging of the resulting magnetic fields generated from the sample for an external field of B$_z=0.316$\,mT (the magnetic field is obtained by fitting the ODMR spectra for each pixel). (d) NV (red) and MOKE (green) images overlapped. }}
\label{nvmoke}
\end{figure*}

In Fig.\, \ref{nvmoke} we show the main result of this work. Figure\,\ref{nvmoke}\,(a) shows the hysteresis magnetization curve of our magnetic sample obtained by implementing a MOKE measurement sequence and by averaging over all pixel values within the FOV for each strength of the externally applied magnetic field (B$_z$) used to generate the magnetization curve [for comparison we also present measurements obtained using a commercial MOKE microscope (Evico GmbH)]. In Fig.\,\ref{nvmoke}\,(b) we show a MOKE image of the sample-magnetization topology under an externally applied magnetic field strength of B$_z=0.316$\,mT, demonstrating, thus, our ability to observe magnetic (stripe) domains and qualitatively image their magnetization. 



In Fig.\,\ref{nvmoke}\,(c) we present the magnetic field map generated by the stripe domains as obtained by NV-based magnetic imaging, together with an example of NV ODMR spectra for two different pixels. One of the pixels above a stripe and one further away. From the difference between the resonance frequencies observed in these ODMR spectra, we obtain the difference in magnetic fields generated by a stripe domain and its surrounding environment, which is $\approx 22\,$\textmu T. As expected, the splitting, between the sets of five lorentzians corresponding to the NV m$_s$\,=\,-1 and m$_s$\,=\,+1 spin states, on top of the stripe domains is reduced compared to the places without stripe domains, since in this regime of the hysteresis loop, most of the spins are still aligned with the externally applied magnetic field while a smaller amount of spins that flipped form the stripe domains. \\
\indent Finally, in Fig.\,\ref{nvmoke}(d) we present the MOKE and NV magnetic images overlapped so that the complementary character of the two imaging modalities becomes clear, while, additionally, we see a difference in the observed size of the stripe domains between the two imaging modalities. This difference is related to an offset distance between diamond and magnetic sample ($\approx2$\,\textmu m, as measured in our microscope) that results in blurring the observed features in the (NV) magnetic image (this also verifies that the resolution of the NV magnetic image is currently determined by the size of this offset distance and not by the optical resolution of the microscope). We note here that this becomes consequential when the stripe domains are dense, i.e. when the distance between the stripe domains is smaller than the resolution of NV-magnetic imaging, resulting in noticeable differences between the NV-magnetic and MOKE images. In such cases, the magnetic fields from different domains might, for example, add up and it will not be easily possible to reconstruct the magnetic topology of the domain structure. However, such an issue can be resolved (in future implementations) by minimizing the distance between the diamond surface and the magnetic sample (for instance, 2D magnetic materials can be directly deposited on the diamond surface\,\cite{Lovchinsky503}). Furthermore, we observe in our images slight deviations from perfect overlap between the observed magnetization and magnetic field maps from stripe features that reside at the edges of our FOV. We believe this is the result of offset-related magnetic field gradients, but further investigation is required to resolve this issue.


\section{Discussion and Conclusions}
We present a novel platform for wide-field imaging of the magnetization and resulting magnetic fields of magnetic structures using engineered diamond magnetic sensors and an optical setup that allows for both measurement modalities. Our work extends recently developed NV-magnetic imaging techniques used to study magnetic systems by demonstrating how the addition of polarization analysis can incorporate simultaneous information about the magnetization of the sample.\\
\indent Possible extensions to the current experimental setup are possible for both the MOKE- and NV-based imaging part. The NV-based imaging, for example, can be extended to the zero-field regime by implementing new measurement schemes\,\cite{Zheng2019} or even be operated in a MW-free modality\,\cite{Zheng2020,Wickenbrock2016}. Furthermore, while the setup is currently only sensitive to out-of-plane magnetizations (polar MOKE), it can be extended to measurements of longitudinal and transverse MOKEs, i.e. the measurement of in-plane magnetizations, by illuminating the sample at an angle \,\cite{McCord2015} and incorporating field coils for the generation of in-plane magnetic fields. Most importantly, the angled illumination allow for truly concurrent acquisition of NV and MOKE images, as the reflected beam and fluorescence are now detected in different areas of the camera (or be independently imaged from their respective optical paths). \\
\indent Novel magnetic structures, and generally correlated and topological electron systems, present crucial technological opportunities for information storage and computation, and their inherent nonvolatility makes them central to energy-related research and associated technologies. Moreover, our understanding of condensed matter physics is greatly affected by these new technologies that exploit the topology of spin structures, as was the case with technologies that exploited the topology of the physical and electronic structure of materials. Our primary goal is to develop a new instrument that is ideal for the research and development of these technologies and the study of related emergent phenomena.


\section{Acknowledgements}
 This work was supported by the EU FETOPEN Flagship Project ASTERIQS (action 820394), the Dynamics and Topology Centre funded by the State of Rhineland Palatinate, the German Federal Ministry of Education and Research (BMBF) within the Quantumtechnologien program (FKZ 13N14439 and FKZ 13N15064), as well as the German Research Foundation (DFG SFB TRB 173 SPIN+X AOI, 403502522). YD acknowledges  support by the Alexander von Humboldt Foundation. K.M.C.F. and Z.K. acknowledge support by the UW Molecular Engineering and Materials Center with funding from the NSF MRSEC program (No.\,DMR-1719797).

\bibliography{NVMOKE}
\end{document}